\documentclass[aps,prl,twocolumn,floats,showpacs]{revtex4}
\usepackage{graphicx}
\usepackage{amsmath}
\usepackage{amssymb}

\begin{document}
\title{Directional spinning of molecules with sequences of femtosecond pulses}

\date{\today}
\author {C.~Bloomquist, S.~Zhdanovich, A.A.~Milner and V.~ Milner}
\affiliation{Department of  Physics \& Astronomy and The Laboratory for Advanced Spectroscopy and Imaging Research (LASIR), The University of British Columbia, Vancouver, Canada \\}
\begin{abstract}{We present an analysis of two experimental approaches to controlling the directionality of molecular rotation with ultrashort laser pulses. The two methods are based on the molecular interaction with either a pair of pulses (a ``double kick'' scheme) or a longer pulse sequence (a ``chiral pulse train'' scheme). In both cases, rotational control is achieved by varying the polarization of and the time delay between the consecutive laser pulses. Using the technique of polarization sensitive resonance-enhanced multi-photon ionization, we show that both methods produce significant rotational directionality. We demonstrate that increasing the number of excitation pulses supplements the ability to control the sense of molecular rotation with quantum state selectivity, i.e. predominant excitation of a single rotational state. We also demonstrate the ability of both techniques to generate counter-rotation of molecular nuclear spin isomers (here, ortho- and para-nitrogen) and molecular isotopologues (here, $^{14}N_{2}$ and $^{15}N_{2}$).}
\end{abstract}


\maketitle

Control of molecular rotation with strong non-resonant laser fields has become a powerful tool for creating ensembles of aligned \cite{Friedrich1995, Stapelfeldt2003, Averbukh2001, Vrakking2001}, oriented \cite{Averbukh2001, Rost1992, Vrakking1997} and spinning molecules \cite{Karczmarek1999, Fleischer2009, Kitano2009, Hoque2011}. Numerous applications of rotational control in molecular systems include control of chemical reactions \cite{Stapelfeldt2003}, deflection of neutral molecules by external fields \cite{Stapelfeldt1997, Purcell2009, Gershnabel2010}, high harmonic generation \cite{Itatani2005, Wagner2007}, and control of molecular collisions with atoms \cite{Tilford2004} and surfaces \cite{ Kuipers1988, Tenner1991, Greeley1995, Zare1998, Shreenivas2010}. Alignment of molecular axes has been implemented with transform-limited and shaped laser pulses using various approaches (see, e.g. \cite{Friedrich1995, Stapelfeldt2003, Underwood05, Lee06, Daems05, Holmegaard09}). Increasing the degree of molecular alignment has been achieved by employing a sequence of laser pulses (a ``pulse train'') separated by the time of rotational revival \cite{Leibscher2003, Cryan2009, Zhao2011, Zhdanovich2012}. Varying the time delay between pulses has been used to promote molecules to a particular angular momentum state \cite{Zhdanovich2011}.

Creating molecular ensembles with a preferred direction of rotation has been demonstrated in the adiabatic regime of excitation with the ``optical centrifuge'' \cite{Karczmarek1999, Villeneuve2000, Vitanov2004, Yuan2011, Cryan2011} and in the non-adiabatic regime of excitation by varying the delay and polarization between a pair of laser pulses (``double kick'') \cite{Fleischer2009, Kitano2009}. In our recent Letter \cite{Zhdanovich2011} we demonstrated an alternative approach to generating uni-directional rotation by means of a``chiral pulse train'' - a sequence of linearly polarized pulses with polarization rotating from pulse to pulse by a controllable angle. By tailoring the train parameters, i.e. the polarization angle and delay between the pulses, we produced clockwise or counter-clockwise uni-directional rotation.

In this work we perform an experimental analysis of both the double-kick and chiral pulse train techniques, and compare their properties. Using the technique of polarization sensitive resonance-enhanced multi-photon ionization (REMPI), we study the degree of rotational directionality induced by the two excitation methods, and its dependence on the excitation parameters. In the case of a chiral pulse train, we show and explain the distinct ability to control the sense of molecular rotation in a state-selective manner, i.e. simultaneously with predominant excitation of a particular Raman transition between angular momentum states. We demonstrate experimentally that both techniques are suitable for creating ensembles of counter-rotating, i.e. rotating in opposite directions, nuclear spin isomers of molecular nitrogen. Producing a sample with oppositely rotating isotopologues, $^{14}N_{2}$ and $^{15}N_{2}$, is also investigated and shown feasible with the double-kick method.

\begin{figure}
\centering
    \includegraphics[width=1.0\columnwidth]{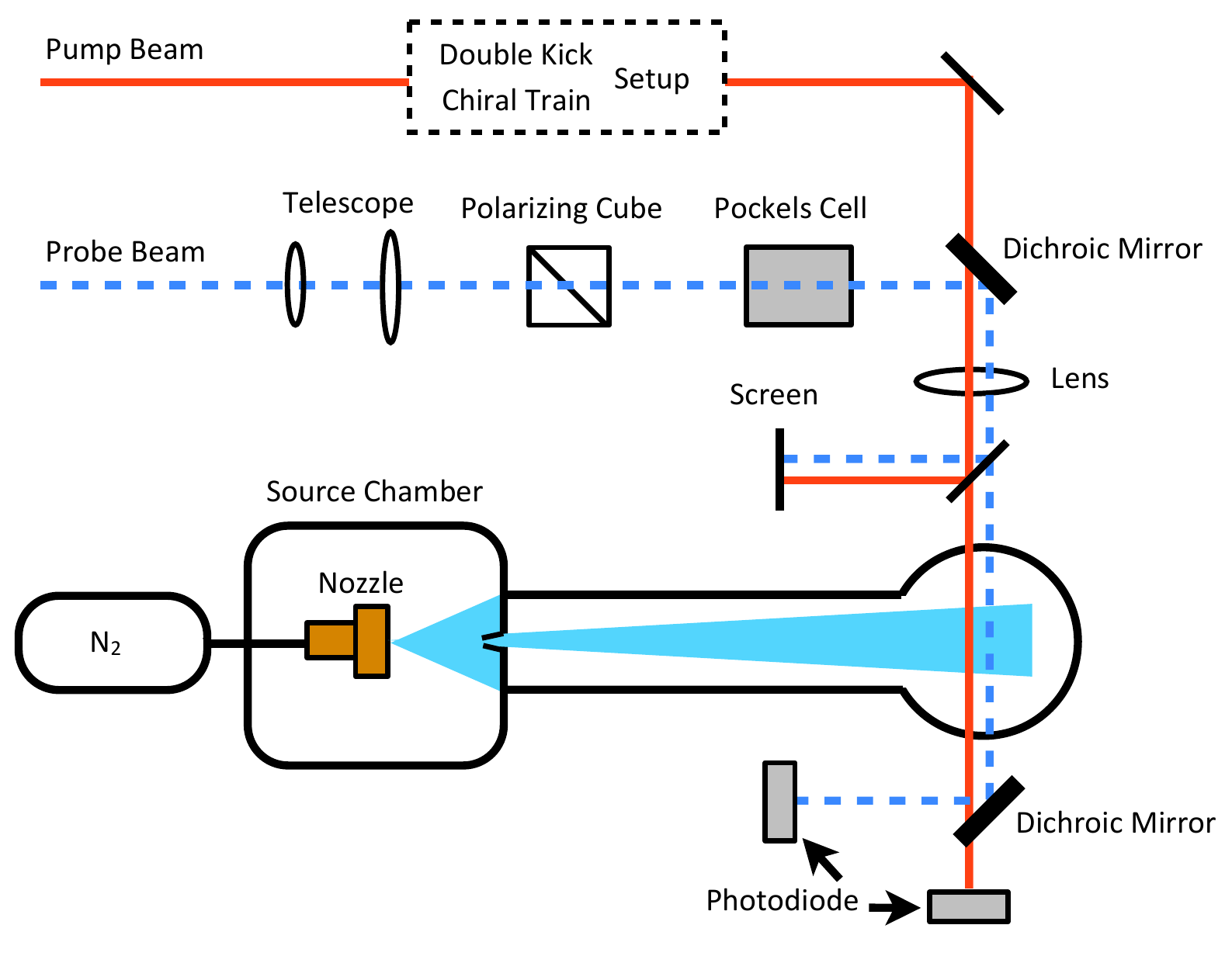}
    \caption{(Color online) Experimental setup. Cold nitrogen molecules from a supersonic expansion enter the detection chamber through a 1 mm-diameter skimmer. The rotational temperature, calculated from the REMPI spectrum, is 6.3 K. The molecules are excited by a double-kick scheme or a chiral pulse train (solid red line) produced by pulse shaping. The probe beam (dashed blue line) is sent through a telescope to control the beam diameter in the focal plane, a polarizing cube to ensure linear polarization, and a Pockels cell to alternate probe polarization between right and left circular. The pump and probe beams are combined on a dichroic mirror and focused on the molecular beam with a 150 mm focal length lens. The pump beam is shifted a few hundred $\mu$m upstream with respect to the probe beam and the ions are extracted and detected with a standard time-of-flight (TOF) apparatus. A paper screen is used to position the beams in space and photodiodes are used to monitor the power of each beam. }
    \label{FigSetup}
\end{figure}

\begin{figure}
\centering
    \includegraphics[width=1.0\columnwidth]{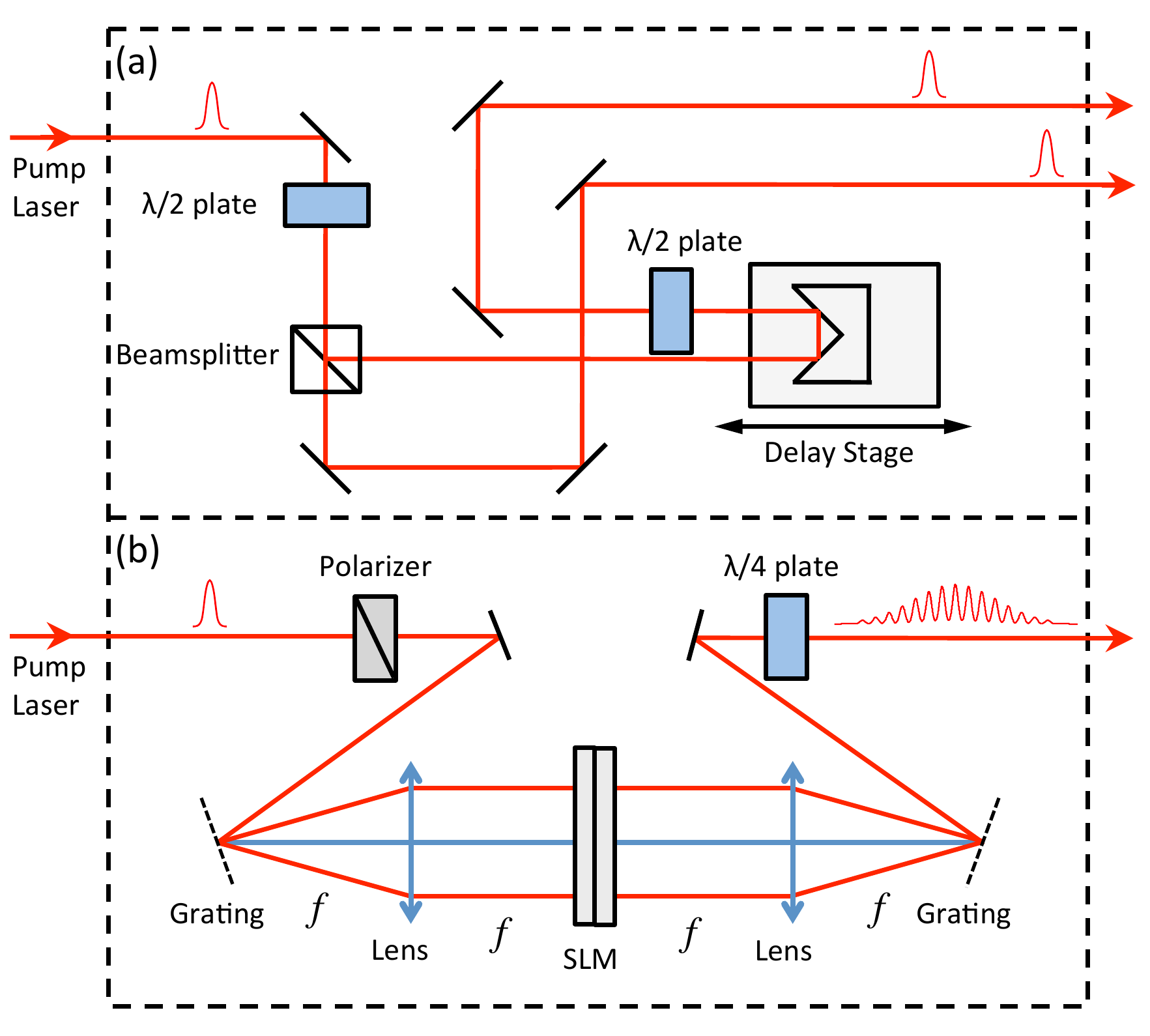}
    \caption{(Color online) Optical setups used for the (a) double-kick and (b) chiral pulse train techniques. The double-kick setup consists of a polarization beam splitter with one arm passing through a computer-controlled delay line and a $\lambda$/2 waveplate for time and polarization control, respectively. The energy ratio of the two pulses is controlled with an input $\lambda$/2  waveplate. The chiral train setup consists of a spectral pulse shaper in a standard $4f$ geometry with a double-layer spatial light modulator (CRi 640) \cite{Weiner2000}. A quarter-wave plate converts elliptical polarization to linear polarization to create the chiral train (see text for details).}
    \label{Figsetup2}
\end{figure}

Our experimental setup is shown in Fig.\ref{FigSetup}. Cold nitrogen is produced by a supersonic expansion from a pulsed valve nozzle (Even-Lavie valve, EL-5-C-S.S.-2010). The rotational temperature of the molecules in the beam is 6-8 K. It is calculated from the detected REMPI spectrum and determines the initial rotational distribution, rather than the rate of collisional decoherence, which is negligible at room temperature and the estimated gas pressure in the beam of $10^-5$ Torr. The molecular beam enters the detection chamber through a 1 mm-diameter skimmer. The molecules are excited by femtosecond pulses from either a beamsplitter with a tunable time delay arm (double-kick), or a home made pulse shaper based on a liquid crystal spatial light modulator (chiral pulse train), as shown in Fig.\ref{Figsetup2}.

The rotational distribution is probed by resonance enhanced multi-photon ionization with narrowband nanosecond pulses from a tunable dye laser (Sirah, Precision Scan, 2 mJ at 283 nm and 10 Hz repetiton rate). The ions are extracted with a time-of-flight (TOF) apparatus and the total ion signal is measured with a microchannel plate detector. In order to reduce the ionization background from pump pulses, the position of the pump beam is shifted by a few hundred $\mu$m upstream with respect to the probe beam, while the time delay between the pulses is set to let the excited molecules reach the probe focal spot. To measure the directionality of molecular rotation with high accuracy and low susceptibility to the power fluctuations of the nanosecond dye laser, a Pockels cell is used to alternate circular polarization of the consecutive probe pulses between left and right. From the symmetry consideration, a molecular state with a certain value $M_J$ of the projection of its total angular momentum on the laser beam axis, is coupled to the ionization continuum by a left circularly polarized probe field with an equal strength to that of a right circular polarization acting on a state with an opposite projection $-M_J$. Hence, the difference in the ionization rate between left and right circularly polarized probe pulses reflects the asymmetry in the $M$-distribution, or equivalently, the directionality of the induced rotation \cite{Kitano2009, Zhdanovich2011}.

Both the double-kick excitation field and the chiral pulse train are obtained from the output of a regenerative amplifier (Spectra-Physics, Spitfire, 120 fs, 2 mJ at 800 nm and 1 KHz repetition rate). In both cases, the time duration of individual pulses in a pulse sequence ($<500$ fs) is much shorter than the rotational period of nitrogen in the low rotational states considered in this work ($J \leq 5$). The double-kick field is obtained using a beamsplitter with a tunable time delay arm as shown in Fig. \ref{Figsetup2}(a). The energy of the two pulses is equalized with an input half-wave plate. One pulse is sent to a stage-mounted retroreflector and a second wave plate to control both the time delay, $\tau$, and polarization angle, $\delta$, relative to the other pulse. The two parallel femtosecond beams are combined with a nanosecond probe beam on a dichroic mirror, and then focused by a 150 mm-focal length lens into the vacuum chamber.

The chiral pulse train is obtained by phase-only shaping by means of a spectral pulse shaper implemented in a standard $4f$ geometry with a double-layer spatial light modulator (SLM) in its Fourier plane as shown in Fig.\ref{Figsetup2}(b) \cite{Weiner2000}. The two shaper masks control the spectral phases $\varphi_{1,2}(\omega )$ of the two polarization components of an input pulse along the two orthogonal axes of the shaper, $\hat{e}_1$ and $\hat{e}_2$. If $\varphi_{1,2}(\omega )=A\sin[(\omega-\omega_{0})\tau +\delta_{1,2}]$, where $\omega_0$ is the optical carrier frequency, $A$ is the modulation amplitude, $\tau $ is the train period and $\delta _{1,2}$ are two arbitrary angles, the resulting field is:
\begin{equation}
\label{EqRotPolPulseTrain}
E(t)=\sum_{i=1,2}{\hat{e}_i (\hat{e}_i \cdot \hat{e}_\text{in})  \sum_{n=-\infty}^{\infty}{J_n(A)\varepsilon(t+n\tau ) \cos[\omega_0t+n\delta_{i}]}},
\end{equation}
where $\varepsilon(t)$ is the electric field envelope of the original pulse polarized along $\hat{e}_\text{in}$. Eq.\ref{EqRotPolPulseTrain} describes a train of elliptically polarized pulses, with the polarization ellipticity of the $n$-th pulse defined by the phase difference $n(\delta _{1}-\delta _{2})$. A quarter-wave plate, oriented along $\hat{e}_\text{in}$, converts this elliptical polarization back to linear, rotated by an angle $n(\delta _{1}-\delta _{2})/2$ with respect to the input polarization. By choosing $\delta _{1}= - \delta _{2} = \delta $, we can create a pulse train with the polarization of each pulse rotated with respect to the previous one by angle $\delta $. The period of the polarization rotation is $T_{p}=2\pi \tau /\delta $.

\begin{figure}
\centering
    \includegraphics[width=1.0\columnwidth]{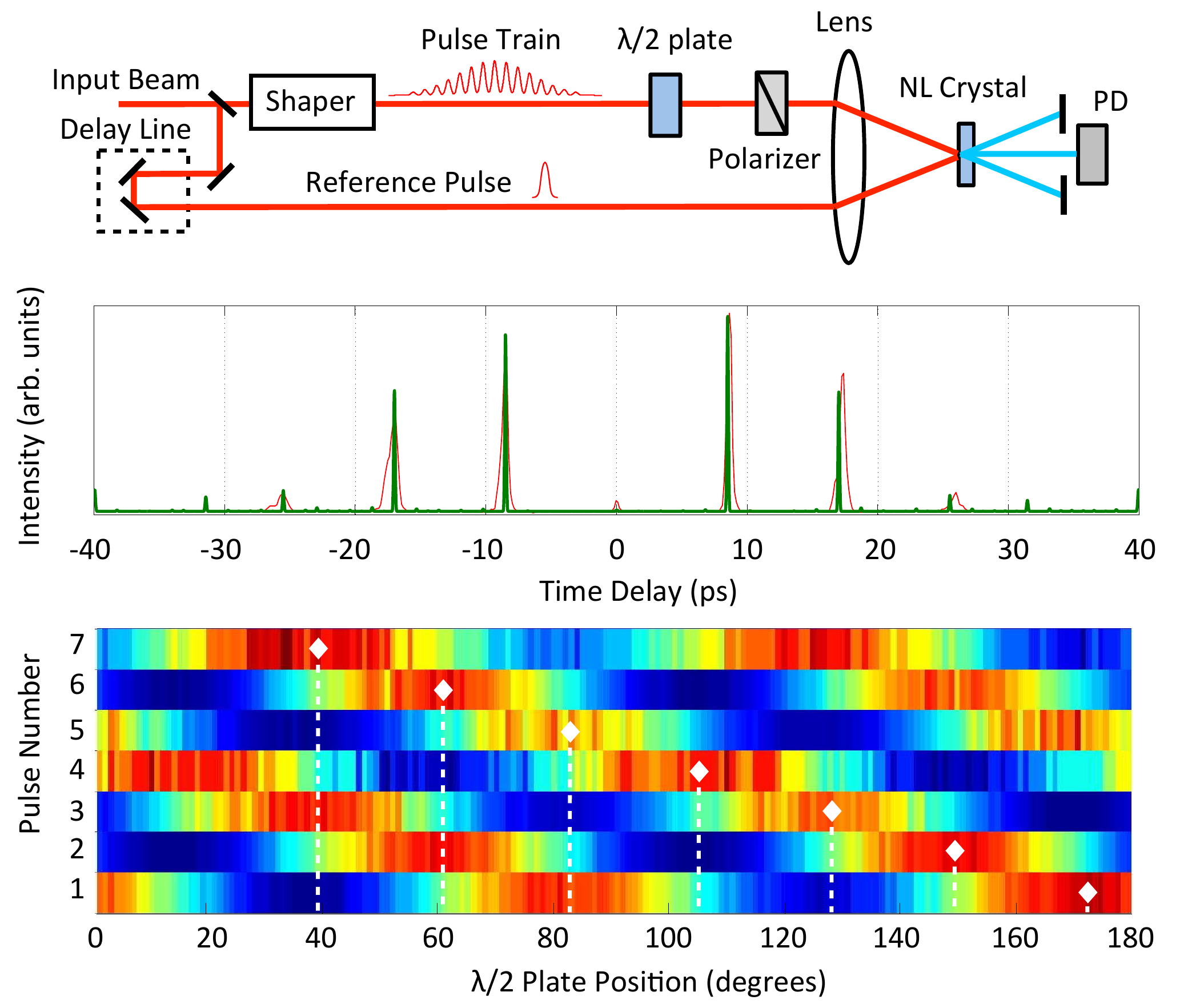}
     \caption{(Color online) Characterization of the chiral pulse train. A polarization sensitive cross-correlation technique is used to characterize the train (upper panel). Middle panel shows the measured (thin red) and calculated (thick green) pulse trains with non-rotating polarization and train period $\tau=8$ ps. Lower panel shows the results of the polarization sensitive cross-correlation measurement for the chiral pulse train. White diamonds and dashed lines mark the peak positions of the non-linear signal for the 7 pulses in the train. The half-wave plate is rotated by $\approx 22.5^{\circ}$ from peak to peak, indicating that the polarization is rotated by $45^{\circ}$ between consecutive pulses ($\delta =\pi/4$).}
  \vskip -.1truein
  \label{Figxfrog}
\end{figure}

\begin{figure}
\centering
    \includegraphics[width=1.0\columnwidth]{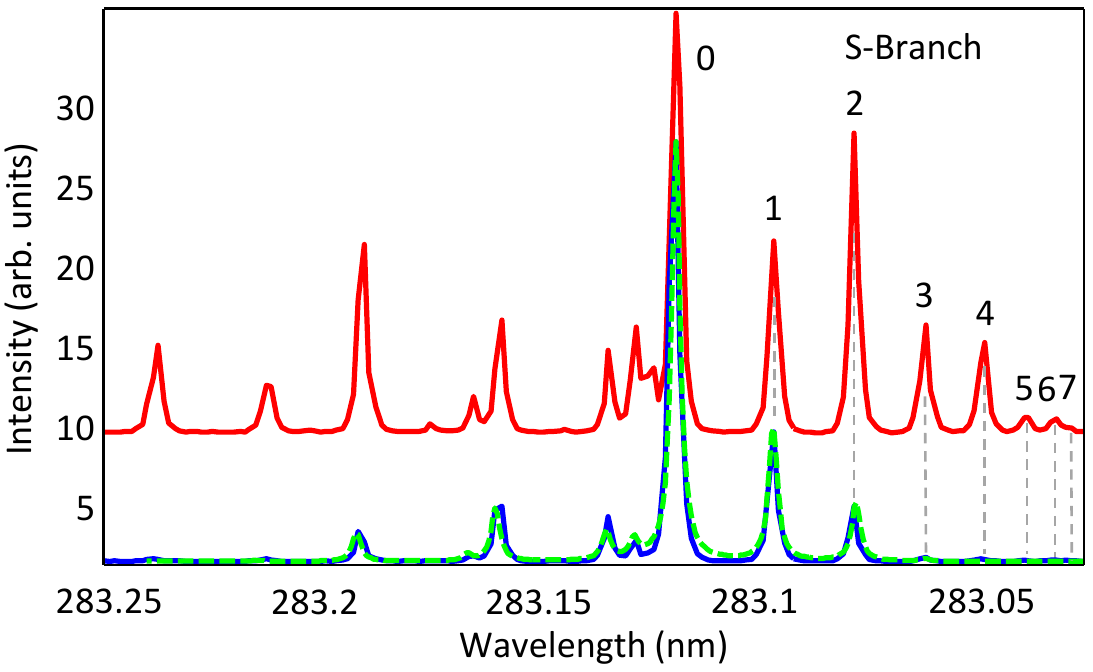}
     \caption{(Color online) REMPI spectra of $ ^{14}N_2$. Relevant peaks of the S-branch are labeled by the corresponding $J''$ numbers. Prior to the application of a femtosecond pulse train, the distribution of rotational population is thermal and corresponds to 6.3K (lower solid blue line - experiment, dashed green line - calculations \cite{HerzbergBook}). At this temperature, only $J''=0, 1$ and 2 are populated significantly. An example of the REMPI spectrum of rotationally excited molecules is shown by upper solid red line. For a total laser kick strength used in our experiments, states up to $J''=7$ are populated.}
  \vskip -.1truein
  \label{FigREMPI}
\end{figure}

We experimentally characterize the chiral pulse train by polarization sensitive cross-correlation technique using the setup shown in Fig.\ref{Figxfrog} (upper panel). Prior to overlapping the train with a transform-limited reference pulse on a nonlinear crystal for second harmonic generation, the train passes through a variable polarization analyzer consisting of a rotatable half-wave plate and a fixed linear polarizer. A delay line is used to overlap the reference pulse with any sub-pulse in the pulse train. Second harmonic signal due to the nonlinear frequency mixing of both input beams is measured as a function of the delay time and the orientation angle of the wave plate.

\begin{figure*}[htp]
\begin{center}
    \begin{minipage}{1\linewidth}
    \includegraphics[width=1\columnwidth]{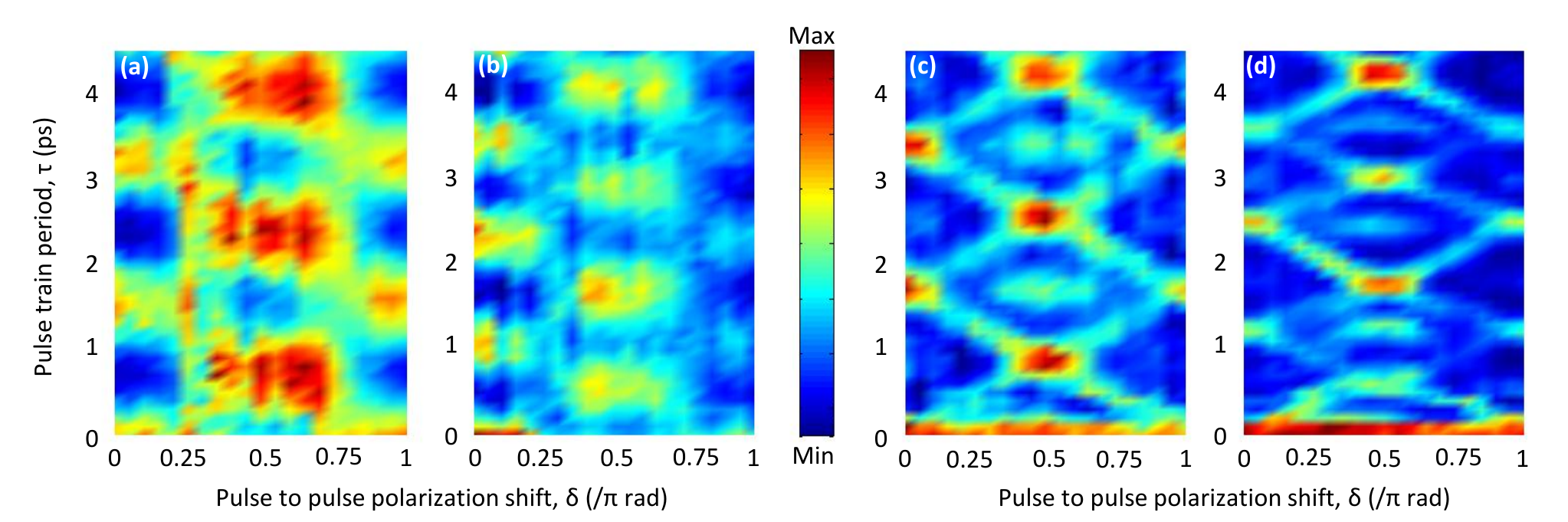}
    \caption{(Color online) Total excitation efficiency for $J''=3$ (a),(c) and $J''=4$ (b),(d) rotational states of nitrogen. (a),(b): double-kick excitation scheme, (c),(d): excitation with the chiral pulse train. In panel (b), the drop-off in signal at higher $\delta$ is caused by a drift of probe wavelength over the course of the scan (2-3 hours). See text for details.}
    \label{2Dex}
    \end{minipage}\hfill
  \end{center}
\end{figure*}

\begin{figure*}
\begin{center}
    \begin{minipage}{1\linewidth}
    \includegraphics[width=1\columnwidth]{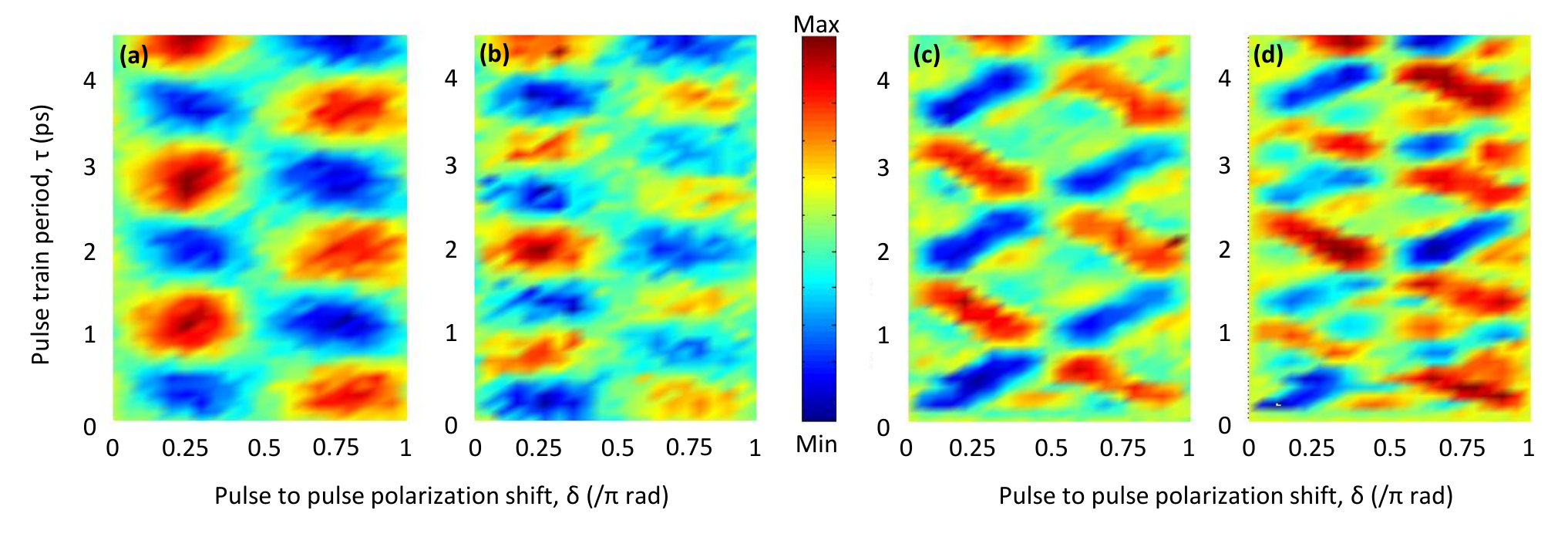}
    \caption{(Color online) Directionality of molecular rotation for $J''=3$ (a),(c) and $J''=4$ (b),(d) rotational states. (a),(b): double-kick excitation scheme, (c),(d): excitation with the chiral pulse train.  The maximum/minimum directionality achieved is $\epsilon=\pm 0.7 $ with the double-kick and $\epsilon=\pm 0.4 $ with the chiral train technique. See text for details.}
    \label{2DCD}
   \end{minipage}\hfill
  \end{center}
\end{figure*}

We first generate a pulse train with train period $\tau=8$ ps and constant polarization ($\delta =0$) coinciding with the polarization of the reference pulse. With the half-wave plate set to have no effect on the train polarization, we observe the cross-correlation trace shown in Fig.\ref{Figxfrog} (middle panel). Then, pulse shaping is applied to generate a chiral train with rotating polarization and the same train period. The time delay of the reference pulse is set to overlap a particular sub-pulse in the pulse train and the cross-correlation signal is measured while the wave plate is rotated. The result for a chiral train with  $\delta=\pi/4$ is shown in Fig.\ref{Figxfrog} (lower panel). Maximizing the cross-correlation signal for each consecutive pulse requires rotating the half-wave plate by $\approx 22.5^{\circ}$, which corresponds to the expected pulse-to-pulse polarization rotation of $45^{\circ}$.

To probe the rotational distribution, we ionize nitrogen molecules via a ``2 +2'' REMPI scheme, with a two-photon resonant transition $a^{1}\Pi _{g}(v'=1) \leftarrow X^{1}\Sigma _{g}^{+}(v''=0)$. The frequencies of the S-branch transitions ($\Delta J=2$) are well separated, allowing us to detect the population of the first eight rotational levels $J''=0,1,...,7$ of the ground electronic state. REMPI spectrum of $^{14}N_{2}$  molecules before and after the application of the excitation laser field is shown in Fig.\ref{FigREMPI}. For most of our measurements we use $J''=3$ and $J''=4$ because these states exhibit the most pronounced change of population as a result of rotational excitation.

To achieve the high laser field intensity required to drive multiple Raman transitions between the rotational levels, pump and probe beams are focused by a 150 mm-focal length lens. After measuring the focal spot size, the peak intensity of unshaped pump pulses was estimated on the order of $10^{13}$ W/cm$^2$. This corresponds to the dimensionless total kick strength $P\approx5$. The latter corresponds to a typical amount of angular momentum (in units of $\hbar$) transferred from the field to the molecule, and is defined as \cite{Averbukh2001,  Leibscher2003}:
\begin{eqnarray}
\label{KSdefenition}
P=\frac{\Delta\alpha}{4\hbar}\int\varepsilon^2dt
\end{eqnarray}
where $\Delta\alpha$ is the anisotropy of the molecular polarizability, and $\varepsilon$ is the electric field amplitude.

Total pulse energy in the double-kick scheme was 500 $\mu $J (250 $\mu $J per pulse), and 190 $\mu $J in the case of the chiral pulse train (the latter being limited by the damage threshold of the pulse shaper). From Eq.\ref{KSdefenition}, one can see that the total kick strength is proportional to the total pulse energy per unit area. However, the different beam sizes in the double-kick and chiral train techniques led to different focal spot sizes, making it difficult to produce the same strength of rotational excitation ($P$) with both techniques. As such, we cannot compare the degree of rotational directionality generated by each technique for the same total kick strength.

We compare the two methods of rotational excitation of nitrogen in Figures \ref{2Dex} and \ref{2DCD}. With the probe wavelength set to 283.06 nm ($J''=3$) and 283.05 nm ($J''=4$), we vary $\delta$ and $\tau$ while keeping the total energy, and hence the total rotational kick strength, constant. We record the ionization signal for left and right circularly polarized probe, $S_{L}$ and $S_{R}$, respectively, and define excitation efficiency as the sum of the two signals, $S=S_L+S_R$. The degree of rotational directionality is defined as $\epsilon  =(S_L-S_R)/(S_L+S_R)$. Figure \ref{2Dex} presents the measured excitation efficiency as a function of $\delta$ and $\tau$, whereas directionality is shown in Figure \ref{2DCD}.

Fig. \ref{2Dex} shows the total excitation efficiency for the double-kick (a,b) and the chiral train (c,d) scheme. For each technique we measure the rotational state population of $J''=3$ (a,c) and $J''=4$ (b,d). Plots corresponding to the same value of $J''$ show maxima in the similar locations (e.g. a maximum at $\delta = \pi/2 , \tau \approx 800$ fs on both Fig.\ref{2Dex} (a) and (c)). For a sequence of pulses with constant polarization $(\delta=0,\pi)$, the total excitation efficiency for $J=3$ and $J''=4$ show maxima at $\tau \approx 1700,3500$ fs and $\tau \approx 1250,2500,3750$ fs, respectively. These times correspond to integer multiples of the ``period of rotation'' for the corresponding rotational states, which refers to the evolution period of a rotational wavepacket consisting of two rotational states separated by $\Delta J=\pm2$, for example $J''=3$ and $J''=1$. Defined as $T_{J} = h/(E_{J}-E_{J-2})$, it results in $T_{J=3}=1677$ fs and $T_{J=4}=1198$ fs.

As shown in our recent Letter \cite{Zhdanovich2011}, higher excitation efficiency signal occurs when the laser polarization rotates ``in-sync'' with the molecules. This effect is demonstrated clearly in the chiral train results (Fig.\ref{2Dex} (c,d)) where the lines of enhanced excitation form a distinct ``X'' pattern. The slope of the lines defines a constant period of polarization rotation in the chiral train, $T_p=2\pi\tau/\delta$. A positive slope corresponds to the polarization rotating clockwise, while a negative slope corresponds to the counter-clockwise rotating polarization. The classical state of enhanced molecular rotation, synchronous with the applied laser field, corresponds to a quantum wavepacket whose amplitude is coherently accumulating with each consecutive laser pulse. With fewer pulses, the excitation efficiency is less sensitive to the train period and polarization angle, resulting in the blurring of the ``X'' pattern in the double-kick results (Fig.\ref{2Dex} (a,b)). For both techniques, the excitation signal is stronger at $\delta=0,\pi/2,\pi$ (middle and end points of the ``X'' pattern) where the conditions of synchronous excitation are satisfied for both directions of rotation.

Fig.\ref{2DCD} shows the degree of rotational directionality, $\epsilon  =(S_L-S_R)/(S_L+S_R)$, measured with the double-kick (a,b) and chiral train (c,d) method. As with excitation efficiency, we measure the rotational state population of $J''=3$ (a,c) and $J''=4$ (b,d) for each technique. No directionality is measured for $\delta=0,\pi$ (non-rotating polarization) or at $\delta=\pi/2$ for which clockwise and counter-clockwise rotating molecules are equally excited, resulting in no preferential sense of rotation in the ensemble. Both methods exhibit the ability to control the direction of molecular rotation. In Fig.\ref{2DCD}(a) and (c) at $\delta=\pi/4, \tau\approx 500$ fs, both techniques produce clockwise rotation ($\epsilon<0$), whereas counter-clockwise rotation ($\epsilon>0$) is observed at $\delta=3\pi/4, \tau\approx 500$ fs. The results for the double-kick scheme show that the strongest directionality is produced when $\delta=\pi/4, 3\pi/4$, in agreement with the calculations of \cite{Fleischer2009}. At these angles, the torque applied to the molecular ensemble by the laser field reaches maximum, resulting in the highest degree of directional excitation. The chiral train also produces the strongest directionality signal when $\delta\approx\pi/4, 3\pi/4$.

The ``X'' patterns of Fig.\ref{2Dex} (c,d) correspond directly to those in Fig.\ref{2DCD} (c,d), confirming the directionality of molecular rotation along the lines of maximum excitation efficiency. As expected, middle and end points of the ``X'' pattern ($\delta=0,\pi/2,\pi$) show $\epsilon  \approx 0$ reflecting bi-directional excitation with no preferential sense of rotation.

\begin{figure}
\centering
    \includegraphics[width=1.0\columnwidth]{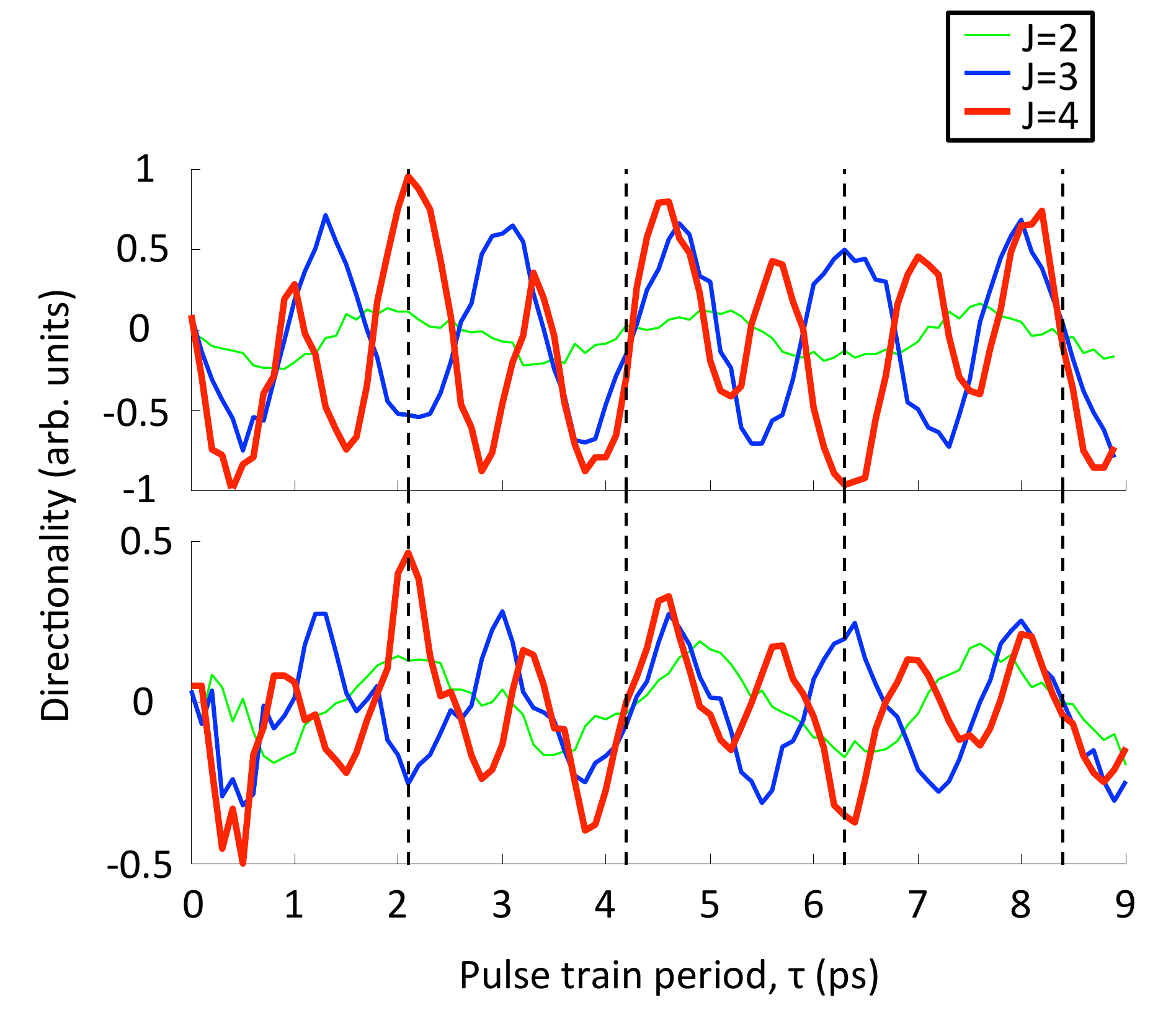}
     \caption{(Color online) Directionality of molecular rotation for $J''=2-4$ with the double-kick (upper panel) and chiral pulse train (lower panel) excitation techniques. In both cases, the relative polarization angle between consecutive pulses is $\delta=\pi/4$. Vertical dashed lines denote $\frac{1}{4} T_{\text{rev}}, \frac{1}{2} T_{\text{rev}}, \frac{3}{4} T_{\text{rev}}$ and full revival time $T_{\text{rev}}$. Higher $J$'s are not shown due to the low signal-to-noise ratio.}
  \vskip -.1truein
  \label{Fig1D}
\end{figure}

\begin{figure*}[t]
\begin{center}
    \begin{minipage}{1\linewidth}
    \includegraphics[width=.7\columnwidth]{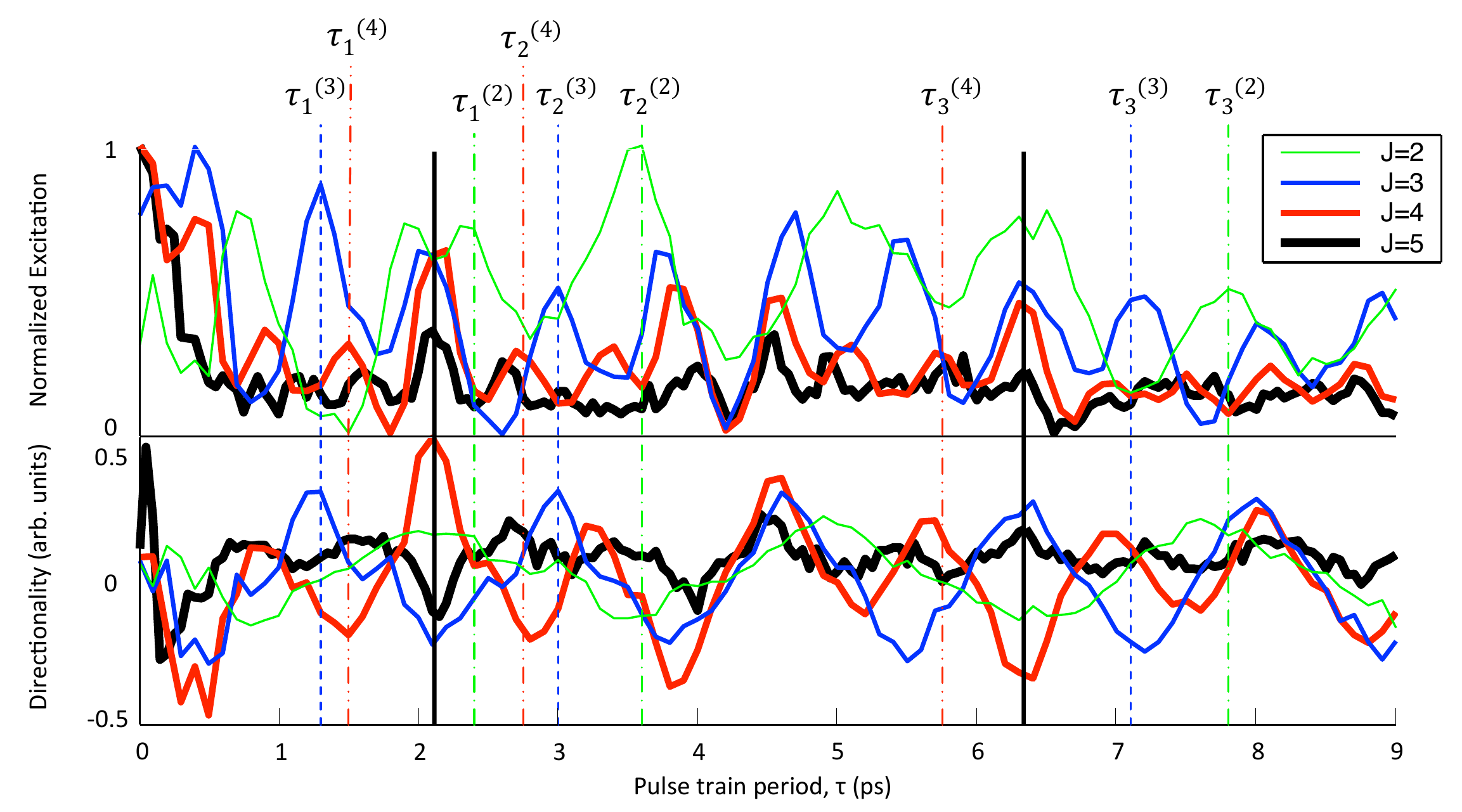}
    \caption{(Color online) Normalized excitation efficiency (a) and degree of directionality (b) for $J=2-5$ states excited with the chiral pulse train of period $\tau $ and pulse-to-pulse polarization angle of $\delta =\pi/4$. Vertical dashed lines indicate pulse periods for selective excitation of a particular rotational state and the corresponding directionality at that time (see text for details). Vertical solid lines mark $\frac{1}{4} T_{\text{rev}}$ and  $\frac{3}{4} T_{\text{rev}}$.}
    \label{Figselectivity}
    \end{minipage}\hfill
  \end{center}
\end{figure*}
To further compare the two techniques, we investigate the degree of rotational directionality as a function of pulse separation $\tau$ at a fixed polarization angle between pulses, $\delta=\pi/4$. Fig.\ref{Fig1D} shows $\epsilon (\tau )$ for  $J''=2-4$ measured with each technique. Both techniques produce very similar structure, particulary at quarter, half, three-quarters and full revival times (vertical dashed lines, $T_{\text{rev}}=8.38$ ps). The behavior around $\frac{1}{2} T_{\text{rev}}$ mirrors the behavior at $T_{\text{rev}}$, with the directionality for all $J$'s going from negative to positive at $\frac{1}{2} T_{\text{rev}}$ and positive to negative at $T_{\text{rev}}$. At $\frac{1}{4} T_{\text{rev}}$ and $\frac{3}{4} T_{\text{rev}}$, the states with even and odd values of angular momentum exhibit opposite sense of rotation. Larger number of pulses in the chiral pulse train (in comparison with the double-kick scheme) leads to higher sensitivity to $\tau $ and correspondingly narrower features in the observed time dependence.

Molecular spin isomers, such as para- and ortho- isomers of $^{14}N_{2}$, exhibit identical revival times yet different structure of rotational levels. Paranitrogen does not have odd $J$ states in its rotational spectrum, whereas even $J$'s are missing in the spectrum of orthonitrogen. After the excitation by a short laser pulse, the alignment factors for spin isomers evolve differently. Around quarter-revival time, they are exactly out of phase, meaning that the ortho molecules are maximally aligned when the para molecules are maximally anti-aligned. Previously, this effect has been used to selectively excite one spin isomer with a sequence of two pulses \cite{Fleischer2007} or a train of equally polarized pulses \cite{Zhdanovich2012}, leaving the other isomer in its ground rotational state. Excitation fields with time-dependent polarization, considered here, produce a very different result. Both the double-kick and chiral pulse train techniques can create a molecular ensemble of oppositely rotating spin isomers. This is shown in Fig.\ref{Fig1D} where at $\tau=\frac{1}{4} T_{\text{rev}}=2100$ fs, positive directionality is detected for even rotational states ($J''=2,4$) and negative directionality - for odd rotational states ($J''=3$).

Notably, the technique of chiral pulse train enables selective directional excitation of not only a set of rotational levels (as in the case with spin isomers described above), but also of individual $J$-states. This is demonstrated in figure \ref{Figselectivity} where we show the normalized excitation signal (a) and the degree of directionality (b) for $J''=2-5$ as a function of the pulse train period $\tau$, at $\delta = \pi/4$. Various dashed lines mark periods that result in a preferential excitation of a particular rotational state and the corresponding directionality at that value of $\tau$. For example, the three train periods that result in the excitation of $J''=3$ while leaving other $J$'s almost unexcited, are labeled $\tau^{(3)}_{1,2,3} \approx 1300, 3000,$ and $7100$ fs. As can be seen in the lower panel of Fig.\ref{Figselectivity}, at $\tau^{(3)}_{1}$ and $\tau^{(3)}_{2}$ positive directionality is measured, indicating counter-clockwise rotation, whereas at $\tau^{(3)}_{3}$ the directionality signal is negative, corresponding to clockwise rotation. Other dashed lines similarly point to values of $\tau $ which provide selective excitation of $J''=2,4$ and 5 in either clockwise or counter-clockwise direction (minimum or maximum in the corresponding directionality signal, respectively). Note that at  $\tau=\frac{1}{4} T_{\text{rev}}$ and  $\tau=\frac{3}{4} T_{\text{rev}}$ (vertical solid lines), all states are excited with maximum efficiency as required for creating an ensemble of counter-rotating isomers discussed above.

The double-kick technique enables an equally efficient control of rotational directionality, but does not offer selectivity in $J$ due to the small number of laser pulses and correspondingly lower time resolution. Inspecting Fig.\ref{2Dex} (a) and (b), one can see that although some limited selectivity of excitation exists at $\delta=0$, it is quite poor at $\delta=\pm\pi/4$ - the angle necessary to induce maximum degree of directionality.

\begin{figure}
\centering
    \includegraphics[width=1.0\columnwidth]{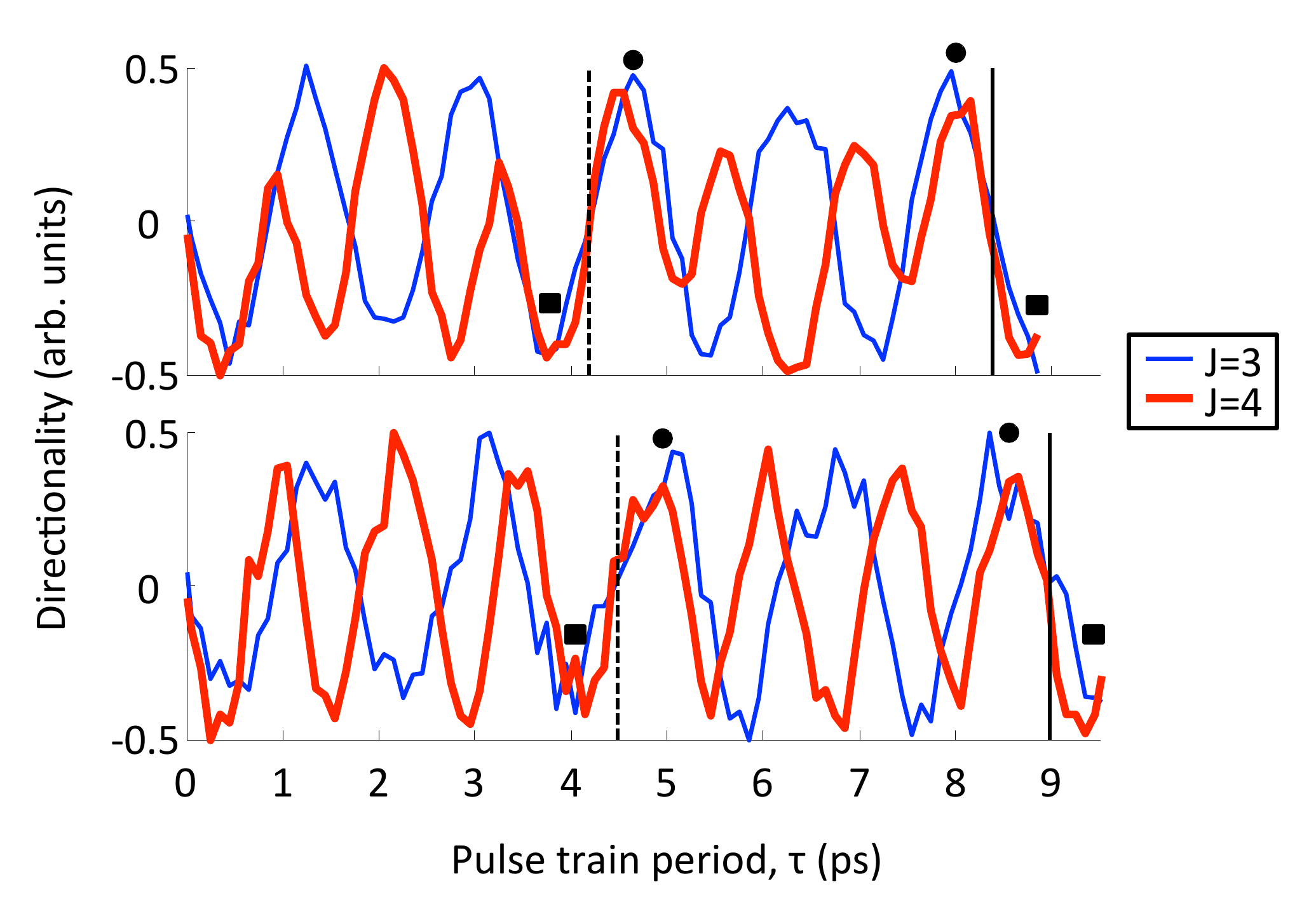}
     \caption{(Color online) Degree of rotational directionality for nitrogen isotopologues, $^{14}N_{2}$ (upper panel) and $^{15}N_{2}$ (lower panel), induced with the double-kick technique at $\delta=\pi/4$. Solid (dashed) lines indicate full (half) revival time for each molecule, $T_{\text{rev}}=8.38$ ps for $^{14}N_{2}$, and $T_{\text{rev}}=8.98$ ps for $^{15}N_{2}$. Black squares (circles) mark the values of $\tau $ resulting in clockwise (counter-clockwise) rotation.}
  \vskip -.1truein
  \label{Figisotope}
\end{figure}

Selective alignment of two molecular isotopologues, $^{14}N_{2}$ and $^{15}N_{2}$, has been first proposed and demonstrated with a sequence of two equally polarized laser pulses \cite{Fleischer2006}. In our recent work \cite{Zhdanovich2012}, we have used the effect of quantum resonance in the periodically kicked rotor system to selectively excite a particular isotopologue of nitrogen with a train of femtosecond pulses of constant polarization. In both approaches, one exploits the fact that different isotopologues exhibit different revival times. Here, we discuss the possibility of creating an ensemble of counter-rotating molecular isotopologues. Fig.\ref{Figisotope} shows the experimentally observed degree of rotational directionality in $^{14}N_{2}$ and $^{15}N_{2}$, excited with the double-kick method. The directionality signal oscillates as a function of $\tau $ with the oscillation period being different for different molecules.

Similarly to the results of Fig.\ref{Fig1D}, the directionality signal for both isotopologues changes from negative (clockwise rotation) to positive (counter-clockwise rotation) at $\frac{1}{2} T_{\text{rev}}$ and vice versa at $T_{\text{rev}}$. Because of the difference in rotational constants, a full revival time of $^{15}N_{2}$ will coincide with a half revival time of $^{14}N_{2}$ at ~70.9 ps \cite{Fleischer2006}. A second laser pulse polarized at an angle of $\delta=\pi/4$ with respect to the first one and applied shortly before or after this characteristic time will produce negative directionality in one isotope and positive directionality in the other. This relatively long time delay can be achieved by the double-kick technique but is beyond the capability of the chiral train method, which is limited by the resolution of a pulse shaper to $\tau \leq 10$ ps.

\begin{figure*}[t]
\begin{center}
    \begin{minipage}{1\linewidth}
    \includegraphics[width=1\columnwidth]{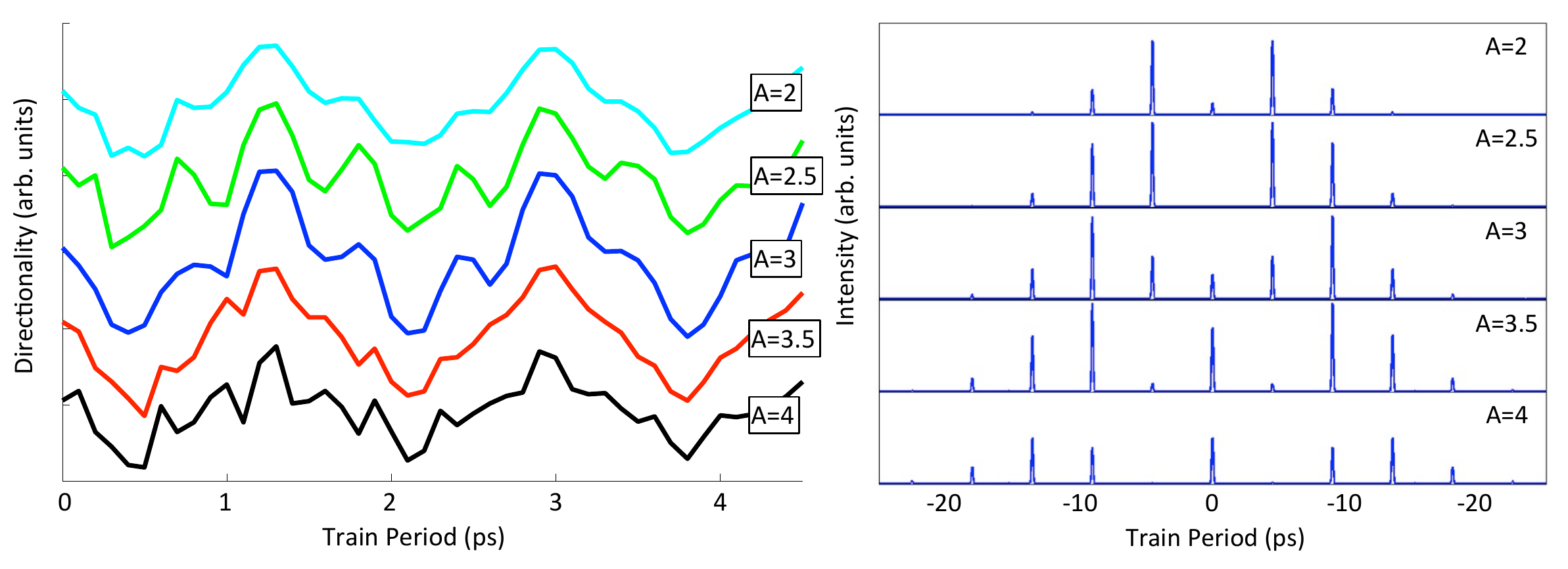}
    \caption{(Color online) Directionality measurements for $J''=3$ and $A=2-4$ (left panel). The plots for each $A$ are offset for clarity, with $A$ increasing from bottom to top. Right panel shows the calculated electric field envelopes for values of the modulation amplitude, $A$, from 2 to 4. The train period is $\tau=4.5$ ps. Polarization rotation that occurs from pulse to pulse in a chiral pulse train is not shown in this figure.}
    \label{FigtrainA3}
    \end{minipage}\hfill
  \end{center}
\end{figure*}

We also explore the effect of different numbers of pulses in the chiral pulse train by varying the modulation amplitude $A$ (see Eq.\ref{EqRotPolPulseTrain}). Left panel of Fig. \ref{FigtrainA3} shows the degree of observed directionality at $J''=3$ with train parameters $\delta=\pi/4$ and variable $A=2.5-4$. Calculated train envelopes for these values of $A$ are shown in the right panel. As the number of pulses in the train increases, the directionality peaks become narrower and new ``side band'' peaks begin to appear. For example, around 1300 fs there are 3 distinct peaks when $A=2.5$ and 5 peaks when $A=4$. This observation agrees well with the theoretical model which shows that the number of local minima between the resonant peaks of $\epsilon (\tau )$ is approximately equal to the number of pulses in the train \cite{Floss2012}.

In summary, we have characterized two experimental techniques for controlling the direction of molecular rotation using sequences of laser pulses with rotating polarization. We have confirmed the ability to control the sense of molecular rotation with both techniques. Using a polarization sensitive and state selective detection technique, we have shown that increasing the number of excitation pulses allows one to control the direction of molecular rotation simultaneously with quantum state selectivity. Using the chiral pulse train method, we have tuned the time delay between pulses and the pulse-to-pulse polarization rotation angle in such a was as to excite a particular rotational state and control the direction of molecular rotation for that state. With both techniques, we have extended previously demonstrated selective excitation of molecular spin isomers and isotopologues by demonstrating the ability to rotate them in opposite directions. This work offers new tools of rotational control in molecular mixtures.

\begin{acknowledgements}
The authors would like to thank Johannes Flo{\ss} and Ilya Sh. Averbukh for valuable discussions.
This work has been supported by the CFI, BCKDF and NSERC. SZ is a recipient of an Alexander Graham Bell scholarship from NSERC.
\end{acknowledgements}


\begin{thebibliography}{39}
\expandafter\ifx\csname natexlab\endcsname\relax\def\natexlab#1{#1}\fi
\expandafter\ifx\csname bibnamefont\endcsname\relax
  \def\bibnamefont#1{#1}\fi
\expandafter\ifx\csname bibfnamefont\endcsname\relax
  \def\bibfnamefont#1{#1}\fi
\expandafter\ifx\csname citenamefont\endcsname\relax
  \def\citenamefont#1{#1}\fi
\expandafter\ifx\csname url\endcsname\relax
  \def\url#1{\texttt{#1}}\fi
\expandafter\ifx\csname urlprefix\endcsname\relax\def\urlprefix{URL }\fi
\providecommand{\bibinfo}[2]{#2}
\providecommand{\eprint}[2][]{\url{#2}}

\bibitem[{\citenamefont{Friedrich and Herschbach}(1995)}]{Friedrich1995}
\bibinfo{author}{\bibfnamefont{B.}~\bibnamefont{Friedrich}} \bibnamefont{and}
  \bibinfo{author}{\bibfnamefont{D.}~\bibnamefont{Herschbach}},
  \bibinfo{journal}{Phys. Rev. Lett.} \textbf{\bibinfo{volume}{74}},
  \bibinfo{pages}{4623} (\bibinfo{year}{1995}).

\bibitem[{\citenamefont{Stapelfeldt and Seideman}(2003)}]{Stapelfeldt2003}
\bibinfo{author}{\bibfnamefont{H.}~\bibnamefont{Stapelfeldt}} \bibnamefont{and}
  \bibinfo{author}{\bibfnamefont{T.}~\bibnamefont{Seideman}},
  \bibinfo{journal}{Rev. Mod. Phys.} \textbf{\bibinfo{volume}{75}},
  \bibinfo{pages}{543} (\bibinfo{year}{2003}).

\bibitem[{\citenamefont{{I. Sh. Averbukh} and Arvieu}(2001)}]{Averbukh2001}
\bibinfo{author}{\bibnamefont{{I. Sh. Averbukh}}} \bibnamefont{and}
  \bibinfo{author}{\bibfnamefont{R.}~\bibnamefont{Arvieu}},
  \bibinfo{journal}{Phys. Rev. Lett.} \textbf{\bibinfo{volume}{87}},
  \bibinfo{pages}{163601} (\bibinfo{year}{2001}).

\bibitem[{\citenamefont{Rosca-Pruna and Vrakking}(2001)}]{Vrakking2001}
\bibinfo{author}{\bibfnamefont{F.}~\bibnamefont{Rosca-Pruna}} \bibnamefont{and}
  \bibinfo{author}{\bibfnamefont{M.~J.~J.} \bibnamefont{Vrakking}},
  \bibinfo{journal}{Phys. Rev. Lett.} \textbf{\bibinfo{volume}{87}},
  \bibinfo{pages}{153902} (\bibinfo{year}{2001}).

\bibitem[{\citenamefont{Rost et~al.}(1992)}]{Rost1992}
\bibinfo{author}{\bibfnamefont{J.~M.} \bibnamefont{Rost}} \bibnamefont{et~al.},
  \bibinfo{journal}{Phys. Rev. Lett.} \textbf{\bibinfo{volume}{68}},
  \bibinfo{pages}{1299} (\bibinfo{year}{1992}).

\bibitem[{\citenamefont{Vrakking and Stolte}(1997)}]{Vrakking1997}
\bibinfo{author}{\bibfnamefont{M.~J.~J.} \bibnamefont{Vrakking}}
  \bibnamefont{and} \bibinfo{author}{\bibfnamefont{S.}~\bibnamefont{Stolte}},
  \bibinfo{journal}{Chem. Phys. Lett.} \textbf{\bibinfo{volume}{271}},
  \bibinfo{pages}{209} (\bibinfo{year}{1997}).

\bibitem[{\citenamefont{Karczmarek et~al.}(1999)}]{Karczmarek1999}
\bibinfo{author}{\bibfnamefont{J.}~\bibnamefont{Karczmarek}}
  \bibnamefont{et~al.}, \bibinfo{journal}{Phys. Rev. Lett.}
  \textbf{\bibinfo{volume}{82}}, \bibinfo{pages}{3420} (\bibinfo{year}{1999}).

\bibitem[{\citenamefont{Fleischer et~al.}(2009)}]{Fleischer2009}
\bibinfo{author}{\bibfnamefont{S.}~\bibnamefont{Fleischer}}
  \bibnamefont{et~al.}, \bibinfo{journal}{New J. Phys.}
  \textbf{\bibinfo{volume}{11}}, \bibinfo{pages}{105039}
  (\bibinfo{year}{2009}).

\bibitem[{\citenamefont{Kitano et~al.}(2009)\citenamefont{Kitano, Hasegawa, and
  Ohshima}}]{Kitano2009}
\bibinfo{author}{\bibfnamefont{K.}~\bibnamefont{Kitano}},
  \bibinfo{author}{\bibfnamefont{H.}~\bibnamefont{Hasegawa}}, \bibnamefont{and}
  \bibinfo{author}{\bibfnamefont{Y.}~\bibnamefont{Ohshima}},
  \bibinfo{journal}{Phys. Rev. Lett.} \textbf{\bibinfo{volume}{103}},
  \bibinfo{pages}{223002} (\bibinfo{year}{2009}).

\bibitem[{\citenamefont{Hoque et~al.}(2011)}]{Hoque2011}
\bibinfo{author}{\bibfnamefont{M.~Z.} \bibnamefont{Hoque}}
  \bibnamefont{et~al.}, \bibinfo{journal}{Phys. Rev. A}
  \textbf{\bibinfo{volume}{84}}, \bibinfo{pages}{013409}
  (\bibinfo{year}{2011}).

\bibitem[{\citenamefont{Stapelfeldt et~al.}(1997)}]{Stapelfeldt1997}
\bibinfo{author}{\bibfnamefont{H.}~\bibnamefont{Stapelfeldt}}
  \bibnamefont{et~al.}, \bibinfo{journal}{Phys. Rev. Lett.}
  \textbf{\bibinfo{volume}{79}}, \bibinfo{pages}{2787} (\bibinfo{year}{1997}).

\bibitem[{\citenamefont{Purcell and Barker}(2009)}]{Purcell2009}
\bibinfo{author}{\bibfnamefont{S.~M.} \bibnamefont{Purcell}} \bibnamefont{and}
  \bibinfo{author}{\bibfnamefont{P.~F.} \bibnamefont{Barker}},
  \bibinfo{journal}{Phys. Rev. Lett.} \textbf{\bibinfo{volume}{103}},
  \bibinfo{pages}{153001} (\bibinfo{year}{2009}).

\bibitem[{\citenamefont{Gershnabel and {I. Sh.
  Averbukh}}(2010)}]{Gershnabel2010}
\bibinfo{author}{\bibfnamefont{E.}~\bibnamefont{Gershnabel}} \bibnamefont{and}
  \bibinfo{author}{\bibnamefont{{I. Sh. Averbukh}}}, \bibinfo{journal}{Phys.
  Rev. Lett.} \textbf{\bibinfo{volume}{104}}, \bibinfo{pages}{153001}
  (\bibinfo{year}{2010}).

\bibitem[{\citenamefont{Itatani et~al.}(2005)}]{Itatani2005}
\bibinfo{author}{\bibfnamefont{J.}~\bibnamefont{Itatani}} \bibnamefont{et~al.},
  \bibinfo{journal}{Phys. Rev. Lett.} \textbf{\bibinfo{volume}{94}},
  \bibinfo{pages}{123902} (\bibinfo{year}{2005}).

\bibitem[{\citenamefont{Wagner et~al.}(2007)}]{Wagner2007}
\bibinfo{author}{\bibfnamefont{N.}~\bibnamefont{Wagner}} \bibnamefont{et~al.},
  \bibinfo{journal}{Phys. Rev. A} \textbf{\bibinfo{volume}{76}},
  \bibinfo{pages}{061403} (\bibinfo{year}{2007}).

\bibitem[{\citenamefont{Tilford et~al.}(2004)}]{Tilford2004}
\bibinfo{author}{\bibfnamefont{K.}~\bibnamefont{Tilford}} \bibnamefont{et~al.},
  \bibinfo{journal}{Phys. Rev. A} \textbf{\bibinfo{volume}{69}},
  \bibinfo{pages}{052705} (\bibinfo{year}{2004}).

\bibitem[{\citenamefont{Kuipers et~al.}(1988)}]{Kuipers1988}
\bibinfo{author}{\bibfnamefont{E.~W.} \bibnamefont{Kuipers}}
  \bibnamefont{et~al.}, \bibinfo{journal}{Nature}
  \textbf{\bibinfo{volume}{334}}, \bibinfo{pages}{420} (\bibinfo{year}{1988}).

\bibitem[{\citenamefont{Tenner et~al.}(1991)}]{Tenner1991}
\bibinfo{author}{\bibfnamefont{M.}~\bibnamefont{Tenner}} \bibnamefont{et~al.},
  \bibinfo{journal}{J. Chem. Phys.} \textbf{\bibinfo{volume}{94}},
  \bibinfo{pages}{5197} (\bibinfo{year}{1991}).

\bibitem[{\citenamefont{Greeley et~al.}(1995)}]{Greeley1995}
\bibinfo{author}{\bibfnamefont{J.~N.} \bibnamefont{Greeley}}
  \bibnamefont{et~al.}, \bibinfo{journal}{J. Chem. Phys.}
  \textbf{\bibinfo{volume}{102}}, \bibinfo{pages}{4996} (\bibinfo{year}{1995}).

\bibitem[{\citenamefont{Zare}(1998)}]{Zare1998}
\bibinfo{author}{\bibfnamefont{R.~N.} \bibnamefont{Zare}},
  \bibinfo{journal}{Science} \textbf{\bibinfo{volume}{279}},
  \bibinfo{pages}{1875} (\bibinfo{year}{1998}).

\bibitem[{\citenamefont{Shreenivas et~al.}(2010)}]{Shreenivas2010}
\bibinfo{author}{\bibfnamefont{D.}~\bibnamefont{Shreenivas}}
  \bibnamefont{et~al.}, \bibinfo{journal}{J. Phys. Chem. A}
  \textbf{\bibinfo{volume}{114}}, \bibinfo{pages}{5674} (\bibinfo{year}{2010}).

\bibitem[{\citenamefont{Underwood et~al.}(2005)\citenamefont{Underwood,
  Sussman, and Stolow}}]{Underwood05}
\bibinfo{author}{\bibfnamefont{J.~G.} \bibnamefont{Underwood}},
  \bibinfo{author}{\bibfnamefont{B.~J.} \bibnamefont{Sussman}},
  \bibnamefont{and} \bibinfo{author}{\bibfnamefont{A.}~\bibnamefont{Stolow}},
  \bibinfo{journal}{Phys. Rev. Lett.} \textbf{\bibinfo{volume}{94}},
  \bibinfo{pages}{143002} (\bibinfo{year}{2005}).

\bibitem[{\citenamefont{Lee et~al.}(2006)}]{Lee06}
\bibinfo{author}{\bibfnamefont{K.~F.} \bibnamefont{Lee}} \bibnamefont{et~al.},
  \bibinfo{journal}{Phys. Rev. Lett.} \textbf{\bibinfo{volume}{97}},
  \bibinfo{pages}{173001} (\bibinfo{year}{2006}).

\bibitem[{\citenamefont{Daems et~al.}(2005)}]{Daems05}
\bibinfo{author}{\bibfnamefont{D.}~\bibnamefont{Daems}} \bibnamefont{et~al.},
  \bibinfo{journal}{Phys. Rev. Lett.} \textbf{\bibinfo{volume}{95}},
  \bibinfo{pages}{063005} (\bibinfo{year}{2005}).

\bibitem[{\citenamefont{Holmegaard et~al.}(2009)}]{Holmegaard09}
\bibinfo{author}{\bibfnamefont{L.}~\bibnamefont{Holmegaard}}
  \bibnamefont{et~al.}, \bibinfo{journal}{Phys. Rev. Lett.}
  \textbf{\bibinfo{volume}{102}}, \bibinfo{pages}{023001}
  (\bibinfo{year}{2009}).

\bibitem[{\citenamefont{Leibscher et~al.}(2003)\citenamefont{Leibscher, {I. Sh.
  Averbukh}, and Rabitz}}]{Leibscher2003}
\bibinfo{author}{\bibfnamefont{M.}~\bibnamefont{Leibscher}},
  \bibinfo{author}{\bibnamefont{{I. Sh. Averbukh}}}, \bibnamefont{and}
  \bibinfo{author}{\bibfnamefont{H.}~\bibnamefont{Rabitz}},
  \bibinfo{journal}{Phys. Rev. Lett.} \textbf{\bibinfo{volume}{90}},
  \bibinfo{pages}{213001} (\bibinfo{year}{2003}).

\bibitem[{\citenamefont{Cryan et~al.}(2009)\citenamefont{Cryan, Bucksbaum, and
  Coffee}}]{Cryan2009}
\bibinfo{author}{\bibfnamefont{J.~P.} \bibnamefont{Cryan}},
  \bibinfo{author}{\bibfnamefont{P.~H.} \bibnamefont{Bucksbaum}},
  \bibnamefont{and} \bibinfo{author}{\bibfnamefont{R.~N.}
  \bibnamefont{Coffee}}, \bibinfo{journal}{Phys. Rev. A}
  \textbf{\bibinfo{volume}{80}}, \bibinfo{pages}{063412}
  (\bibinfo{year}{2009}).

\bibitem[{\citenamefont{Zhao et~al.}(2011)}]{Zhao2011}
\bibinfo{author}{\bibfnamefont{S.}~\bibnamefont{Zhao}} \bibnamefont{et~al.},
  \bibinfo{journal}{Chem. Phys. Lett.} \textbf{\bibinfo{volume}{506}},
  \bibinfo{pages}{26} (\bibinfo{year}{2011}).

\bibitem[{\citenamefont{Zhdanovich et~al.}(2012)}]{Zhdanovich2012}
\bibinfo{author}{\bibfnamefont{S.}~\bibnamefont{Zhdanovich}}
  \bibnamefont{et~al.}, \bibinfo{journal}{Phys. Rev. Lett.}
  \textbf{\bibinfo{volume}{109}}, \bibinfo{pages}{043003}
  (\bibinfo{year}{2012}).

\bibitem[{\citenamefont{Zhdanovich et~al.}(2011)}]{Zhdanovich2011}
\bibinfo{author}{\bibfnamefont{S.}~\bibnamefont{Zhdanovich}}
  \bibnamefont{et~al.}, \bibinfo{journal}{Phys. Rev. Lett.}
  \textbf{\bibinfo{volume}{107}}, \bibinfo{pages}{243004}
  (\bibinfo{year}{2011}).

\bibitem[{\citenamefont{Villeneuve et~al.}(2000)}]{Villeneuve2000}
\bibinfo{author}{\bibfnamefont{D.~M.} \bibnamefont{Villeneuve}}
  \bibnamefont{et~al.}, \bibinfo{journal}{Phys. Rev. Lett.}
  \textbf{\bibinfo{volume}{85}}, \bibinfo{pages}{542} (\bibinfo{year}{2000}).

\bibitem[{\citenamefont{Vitanov and Girard}(2004)}]{Vitanov2004}
\bibinfo{author}{\bibfnamefont{N.~V.} \bibnamefont{Vitanov}} \bibnamefont{and}
  \bibinfo{author}{\bibfnamefont{B.}~\bibnamefont{Girard}},
  \bibinfo{journal}{Phys. Rev. A} \textbf{\bibinfo{volume}{69}},
  \bibinfo{pages}{033409} (\bibinfo{year}{2004}).

\bibitem[{\citenamefont{Yuan et~al.}(2011)}]{Yuan2011}
\bibinfo{author}{\bibfnamefont{L.}~\bibnamefont{Yuan}} \bibnamefont{et~al.},
  \bibinfo{journal}{PNAS} \textbf{\bibinfo{volume}{108}}, \bibinfo{pages}{6872}
  (\bibinfo{year}{2011}).

\bibitem[{\citenamefont{Cryan et~al.}(2011)}]{Cryan2011}
\bibinfo{author}{\bibfnamefont{J.~P.} \bibnamefont{Cryan}}
  \bibnamefont{et~al.}, \bibinfo{journal}{Phys. Rev. X}
  \textbf{\bibinfo{volume}{1}}, \bibinfo{pages}{011002} (\bibinfo{year}{2011}).

\bibitem[{\citenamefont{Weiner}(2000)}]{Weiner2000}
\bibinfo{author}{\bibfnamefont{A.~M.} \bibnamefont{Weiner}},
  \bibinfo{journal}{Rev. Sci. Instrum.} \textbf{\bibinfo{volume}{71}},
  \bibinfo{pages}{1929} (\bibinfo{year}{2000}).

\bibitem[{\citenamefont{Herzberg}(1950)}]{HerzbergBook}
\bibinfo{author}{\bibfnamefont{G.}~\bibnamefont{Herzberg}},
  \emph{\bibinfo{title}{Molecular spectra and molecular structure}},
  vol.~\bibinfo{volume}{I} (\bibinfo{publisher}{Krieger publishing co},
  \bibinfo{address}{Malabar, Florida}, \bibinfo{year}{1950}),
  \bibinfo{edition}{2nd} ed.

\bibitem[{\citenamefont{Fleischer et~al.}(2007)\citenamefont{Fleischer, {I. Sh.
  Averbukh}, and Prior}}]{Fleischer2007}
\bibinfo{author}{\bibfnamefont{S.}~\bibnamefont{Fleischer}},
  \bibinfo{author}{\bibnamefont{{I. Sh. Averbukh}}}, \bibnamefont{and}
  \bibinfo{author}{\bibfnamefont{Y.}~\bibnamefont{Prior}},
  \bibinfo{journal}{Phys. Rev. Lett.} \textbf{\bibinfo{volume}{99}},
  \bibinfo{pages}{093002} (\bibinfo{year}{2007}).

\bibitem[{\citenamefont{Fleischer et~al.}(2006)\citenamefont{Fleischer, {I. Sh.
  Averbukh}, and Prior}}]{Fleischer2006}
\bibinfo{author}{\bibfnamefont{S.}~\bibnamefont{Fleischer}},
  \bibinfo{author}{\bibnamefont{{I. Sh. Averbukh}}}, \bibnamefont{and}
  \bibinfo{author}{\bibfnamefont{Y.}~\bibnamefont{Prior}},
  \bibinfo{journal}{Phys. Rev. A} \textbf{\bibinfo{volume}{74}},
  \bibinfo{pages}{041403} (\bibinfo{year}{2006}).

\bibitem[{\citenamefont{Flo{\ss} and {I. Sh. Averbukh}}(2012)}]{Floss2012}
\bibinfo{author}{\bibfnamefont{J.}~\bibnamefont{Flo{\ss}}} \bibnamefont{and}
  \bibinfo{author}{\bibnamefont{{I. Sh. Averbukh}}}, \bibinfo{journal}{In
  preparation}  (\bibinfo{year}{2012}).

\end{thebibliography}

\end{document}